\documentclass[superscriptaddress,twocolumn,showpacs,showkeys,amsmath,aps,prb]{revtex4-1}

\usepackage{graphicx}
\usepackage{dcolumn}
\usepackage[sort&compress]{natbib}
\usepackage{subfigure}
\usepackage{bm}
\usepackage{ifpdf}

\ifpdf
\usepackage[pdftex,
        colorlinks=true,
        pdftitle={Transverse magnetization and torque in asymmetrical mesoscopic superconductors},
        pdfauthor={Mauro M. Doria},
        pdfsubject={Little-Parks Oscillations},
        pdfkeywords={GL theory, Comsol, Persistent current},
        baseurl={http://www.if.ufrj.br/~mmd}
        pdfpagemode=UseOutlines,
        pdfstartview=FitH,
        bookmarksopen=true
        pdfoutput=1
        ]{hyperref}
\fi

\begin{document}

\title{Little-Parks oscillations near a persistent current loop}
\author{Rodolpho Ribeiro Gomes}%
\affiliation{Centro Brasileiro de Pesquisas F\'{\i}sicas,  22290-160 Rio de Janeiro RJ Brazil}%
\author{Isa\'{\i}as G. de Oliveira}%
\affiliation{Departamento de F\'{\i}sica, Universidade Federal Rural do Rio de Janeiro, 23890-000 Serop\'edica RJ Brazil}%
\author{Mauro M. Doria}%
\affiliation{Departamento de F\'{\i}sica dos S\'{o}lidos, Universidade Federal do Rio de Janeiro, 21941-972 Rio de Janeiro RJ, Brazil}%
\email{mmd@if.ufrj.br}

\begin{abstract}
We investigate the Little-Parks oscillations caused by a persistent
current loop set on the top edge of a mesoscopic superconducting
thin-walled cylinder with a finite height. For a short cylinder the
Little-Parks oscillations are approximately the same ones of the
standard effect, as there is only one magnetic flux piercing the
cylinder. For a tall cylinder the inhomogeneity of the magnetic
field makes different magnetic fluxes pierce the cylinder at
distinct heights and we show here that this produces two distinct
Little-Parks oscillatory regimes according to the persistent current
loop. We show that these two regimes, and also the transition
between them, are observable in current measurements done in the
superconducting cylinder. The two regimes stem from different
behavior along the height, as seen in the order parameter,
numerically obtained from the Ginzburg-Landau theory through the
finite element method.
\end{abstract}

\maketitle
\section{Introduction}\label{Introduction}
In the last decades we have witnessed a huge progress in the
microfabrication of systems and measurement techniques that have
allowed the study of mesoscopic superconducting
structures~\cite{moshchalkov93}. The critical temperature of a
superconducting ring oscillates according to the applied external
magnetic field, as shown by Little and Parks (LP) in
1962~\cite{lp62}. The LP effect can be regarded as a forerunner of
recent developments~\cite{moshchalkov95}, because of its eminently
mesoscopic nature. It predicts that for a thin-walled cylinder of
radius $R$, the critical temperature varies as $\Delta T_c =
(\hbar^2/8 m R^2)(\Phi/\Phi_0-n)^2$, where m is the Copper pair
mass, $\Phi_0$ is the fundamental flux, $\Phi=H \,\pi R^2$ is the
magnetic flux trapped inside the ring, and $n$ describes a quantum
number. A temperature variation detectable within experimental
range, say $\Delta Tc \sim 10^{-5}$ K, means that the radius must be
in the mesoscopic domain, $R \sim 1.0 \mu$m. The LP oscillations are
detected by measuring two consecutive temperature maxima, each
occurring for $\Delta T_c =0$, when the total magnetic flux $\Phi$,
which is the sum of the external magnetic flux piercing a given
surface and the magnetic flux produced by the circulation of the
screening supercurrents along the curve bounding this surface, add
up to $n\Phi_0$. Since its initial proposal the LP effect has been
measured by several different techniques and several distinct
systems, such as a perforated disk with varying hole
size~\cite{morelle04}, a patterned microstructure of an oxide
superconductor~\cite{gammel90}, a single mesoscopic Al ring (SQUID
measurement of the susceptibility)~\cite{zhang97}, an array of Al
loops (specific heat measurement)~\cite{bourgeois05}, and  a single
YBCO sub-micron ring~\cite{carillo10}. The LP effect can be used to
demonstrate many features of the superconducting state, such as the
interaction among two superconducting order
parameters~\cite{bluhm06,erin08} and the study of
superconductor/ferromagnet hybrids~\cite{aladyshkin07}. The LP
effect is well described by the Ginzburg-Landau theory in most of
the situations~\cite{berger95,benoist97,bruyndoncx99,zha11}. The LP
can also exist through fluctuations in the normal phase for very
small rings with effective radius $R/\xi<0.6$ under a constant
magnetic field~\cite{schwiete10}.

In this paper we investigate the LP effect in a thin-walled cylinder
with a finite height produced by the inhomogeneous magnetic field of
a persistent current loop, put on its top. This external magnetic
field is that of a magnetic moment pointing along the thin-walled
cylinder major axis. In the standard LP effect the height of the
thin-walled cylinder plays no role because the applied external
field is constant and oriented along the major axis. However in the
presence situation the height plays a major
role~\cite{aladyshkin07}. We report here new features due to the
inhomogeneity distribution of the supercurrent in the cylinder,
which in its top edge feels a magnetic flux more intense than in the
bottom. Our theoretical framework is developed in a current bias
regime, which means that the persistent current circulating in the
loop is varied and the response of the thin-walled cylinder
observed.

\begin{figure}[b]
\includegraphics[width=0.75\linewidth]{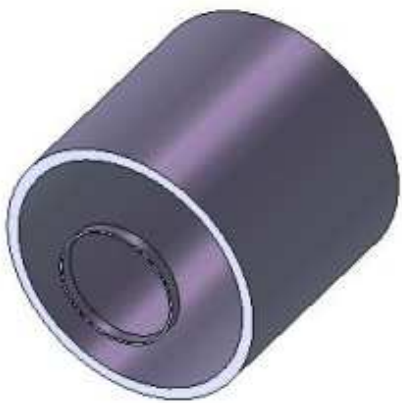}
\caption{(Color online) %
This figure depicts the persistent current loop (radius
$a=5.0\xi_0$) on top of a thin-walled superconducting rod (radius
$R=10.0\xi_0$) put $0.5\xi_0$ away from its edge. Two heights are
treated here, corresponding to a short ($Z_0=2.0\xi_0$) and a tall
($10.0\xi_0$)cylinder.}\label{draw}
\end{figure}

The persistent current loop  of a (non-superconducting) metallic
mesoscopic ring has been observed since decades ago~\cite{levy90}
and in principle can be used for an experimental realization of the
present proposal. The persistent current loop created by a
superconducting mesoscopic ring is another way to realize the
present system. In this case the system has two superconductors, for
instance, Nb and Al, the former used for the persistent current loop
and the latter for the the thin-walled cylinder. Since the Nb
critical temperature (T$_c$ = 9.25 K) is 7.8 times larger than that
of Al (T$_c$ = 1.18 K)~\cite{orlando91}, the LP oscillations
observed in the thin-walled cylinder are near to the Al transition
to the normal state, which happens to fall in the very low
temperature domain of the Nb current loop. Therefore the Nb ring
generates a stable steady magnetic whose value is swept continuously
from zero to its maximum value by varying the persistent current to
the maximum Nb critical current. Let $a$ and $R$ be the radii of the
external current loop and of the thin-walled cylinder, respectively,
as shown in Fig.~\ref{draw}. We are mostly interested here in the
effects brought by the inhomogeneity of the field, and for this
reason will concentrate in the case $R > a$. The magnetic moment of
the persistent current loop is given by $\mu=I_{loop}\,\pi a^2$ but,
notice that the thin-walled cylinder also defines a magnetic moment
scale, given by $\mu_0 = \Phi_0\xi_0/2\pi$, where $\xi_0$ is the
zero temperature coherence length of the superconductor that makes
the Al thin-walled cylinder. We study the LP oscillations by
sweeping the ratio $\mu/\mu_0$ to its maximum value $\mu_c/\mu_0$.
Thus it is of fundamental importance to determine the maximum
attained magnetic moment in the current loop, $\mu_c=I_c\,\pi a^2$,
$I_c$ being the Nb critical current. According to Sebastien Michotte
et al.~\cite{michotte10} for T= 1.5 K  the Nb critical current is
I$_c$=2000 $\mu$A for a wire of width 75 nm. Therefore a ring with
radius a=0.5 $\mu$m achieves the maximum magnetic moment $\mu_c$=1.5
x 10$^{-16}$ A.m$^2$. To determine the magnetic moment scale $\mu_0$
of the thin-walled cylinder it suffices to know the zero temperature
coherence length of Al. Nanoscopic Al rings have been extensively
studied in the literature, such as by Florian R. Ong et
al.~\cite{bourgeois05} and Hendrik Bluhm et al.~\cite{bluhm06}. Data
fit in both cases leads to the values of $\xi_0$ equal to 150 and 70
nm, respectively. Here we shall consider $\xi_0$=100 nm which gives
$\mu_0$=3.2 x 10$^{-23}$. In summary we find that maximum magnetic
moment ratio is $\mu_c/\mu_0$=3 x 10$^{6}$. Therefore  the current
circulating in the external loop can be swept up to six orders of
magnitude in our suggested system. Notice that our theoretical
calculations do not rely in this specific realization which is only
a suggested experimental set up for it.

\begin{figure}[b]
\includegraphics[width=1.00\linewidth]{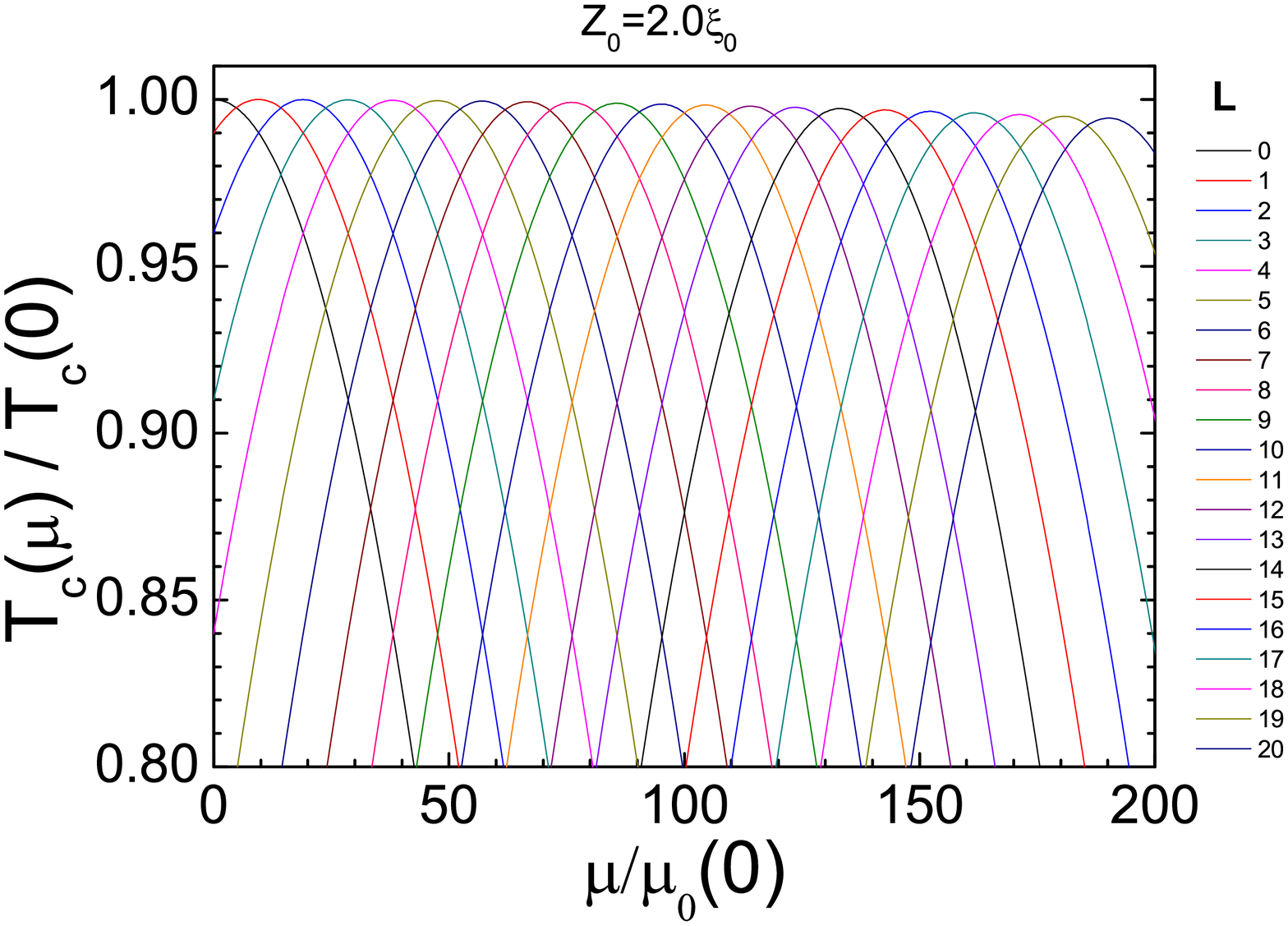}
\caption{(Color online) %
The LP temperature $T_c(\mu)$ versus the magnetic moment $\mu$ is
shown here for the short cylinder (height $Z_0=2.0\xi_0$). The $L$
lines, ranging from $0$ to $20$, are obtained from the linear theory
(Eq.(\ref{linearf})) and describe angular momenta trapped in the
cylinder.}\label{lin_z2}
\end{figure}

\begin{figure}[b]
\includegraphics[width=1.0\linewidth]{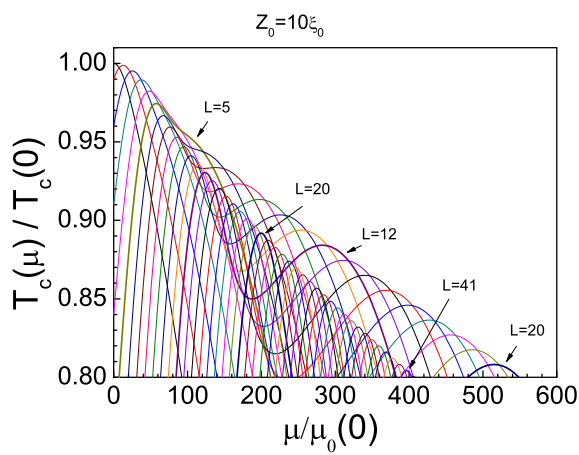}
\caption{(Color online) %
The LP temperature $T_c(\mu)$ versus the magnetic moment $\mu$ is
shown here for the tall cylinder (height $Z_0=10.0\xi_0$). The $L$
lines, ranging from $0$ to $41$, are obtained from the linear theory
(Eq.(\ref{linearf})) and describe angular momenta states trapped in
the cylinder. A few selected curves are shown as thick lines and
their properties discussed in the text.}\label{lin_z10}
\end{figure}

\begin{figure}[b]
\includegraphics[width=1.0\linewidth]{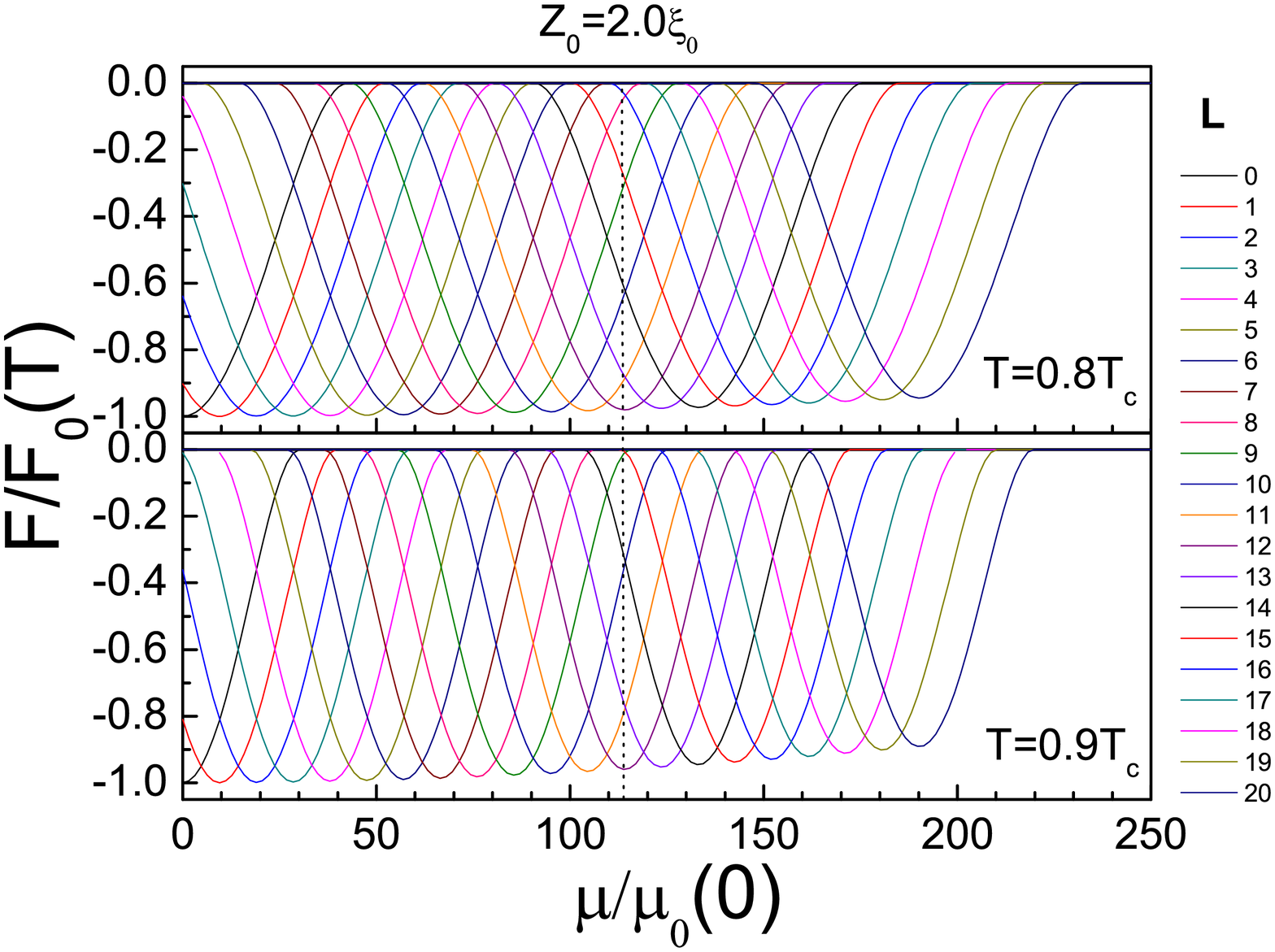}
\caption{(Color online) %
The free energy of the short cylinder (height $Z_0=2.0\xi_0$) is
shown here for two temperatures $T=0.8T_c$ and $T=0.9T_c$. Free
energy $L$ lines ranging from $0$ to $20$ are shown versus the
magnetic moment $\mu$ of the persistent current loop. A dashed
vertical line indicates that the position of the minima is
approximately the same for the two temperatures.
}\label{free_z2_t08t09}
\end{figure}

\begin{figure}[b]
\includegraphics[width=1.0\linewidth]{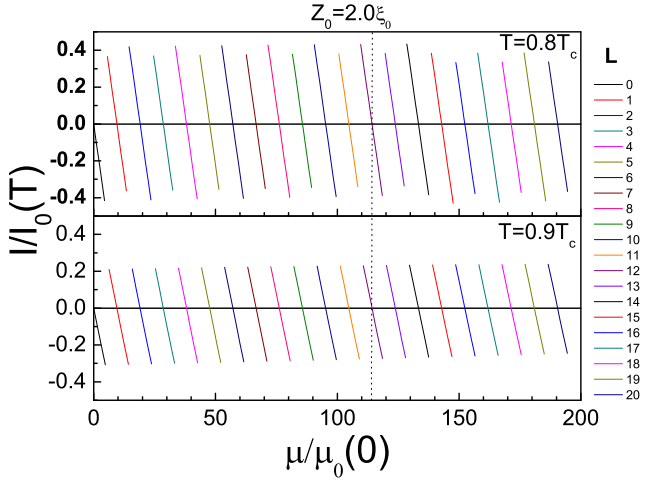}
\caption{(Color online) %
The current circulating in the short cylinder (height
$Z_0=2.0\xi_0$) is shown here for two temperatures $T=0.8T_c$ and
$T=0.9T_c$. Current $L$ lines ranging from $0$ to $20$ are shown
versus the magnetic moment $\mu$ of the persistent current loop. A
dashed vertical line indicates that the vanishing of the current is
approximately the same for the two temperatures.
}\label{curr_z2_t08t09}
\end{figure}

\begin{figure}[b]
\includegraphics[width=1.0\linewidth]{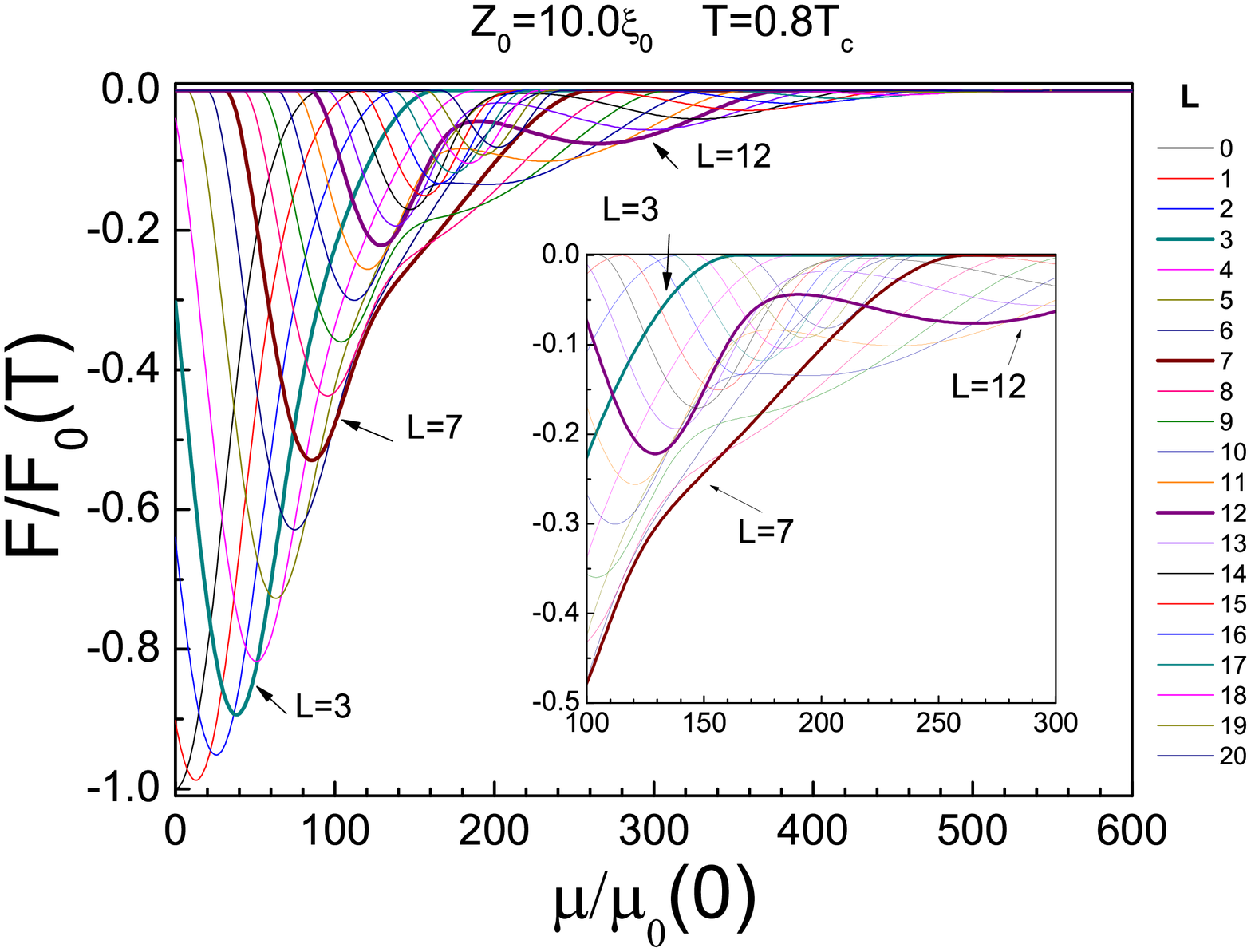}
\caption{(Color online) %
The free energy of the tall cylinder (height $Z_0=10.0\xi_0$) is
shown here for the temperature $T=0.8T_c$. Free energy $L$ lines
ranging from $0$ to $20$ are shown versus the magnetic moment $\mu$
of the persistent current loop. Some $L$ states are depicted as
thick lines to exemplify the three existing regimes: single well
($L=3$), transition ($L=7$), and double well ($L=12$). The inset
helps to visualize the transition regime. }\label{free_z10_t08}
\end{figure}

\begin{figure}[b]
\includegraphics[width=1.0\linewidth]{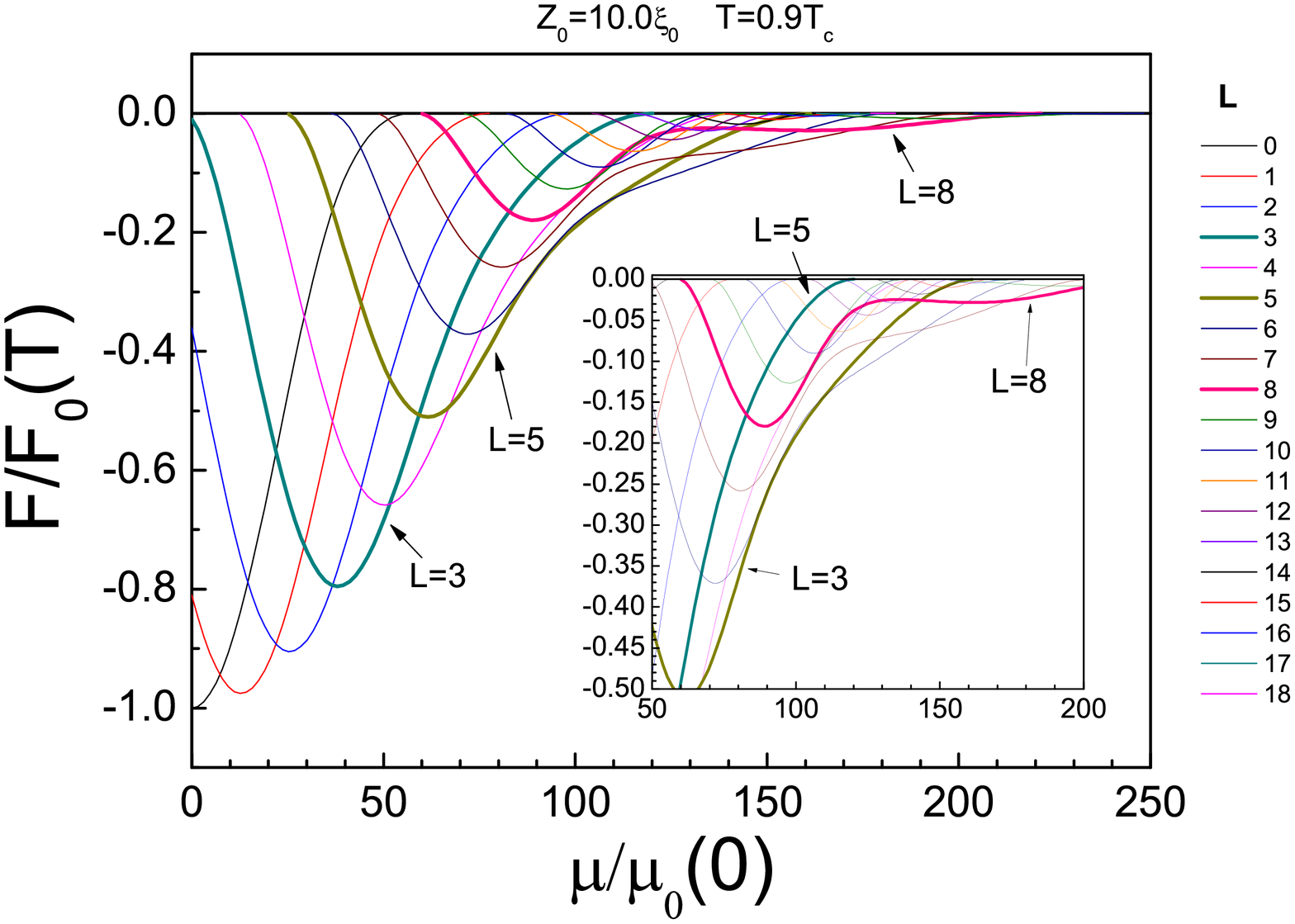}
\caption{(Color online) %
The free energy of the tall cylinder (height $Z_0=10.0\xi_0$) is
shown here for the temperature $T=0.9T_c$. Free energy $L$ lines
ranging from $0$ to $18$ are shown versus the magnetic moment $\mu$
of the persistent current loop. Some $L$ states are depicted as
thick lines to exemplify the three existing regimes: single well
($L=3$), transition ($L=5$), and double well ($L=8$). The inset
helps to visualize the transition regime.}\label{free_z10_t09}
\end{figure}

\begin{figure}[b]
\includegraphics[width=1.0\linewidth]{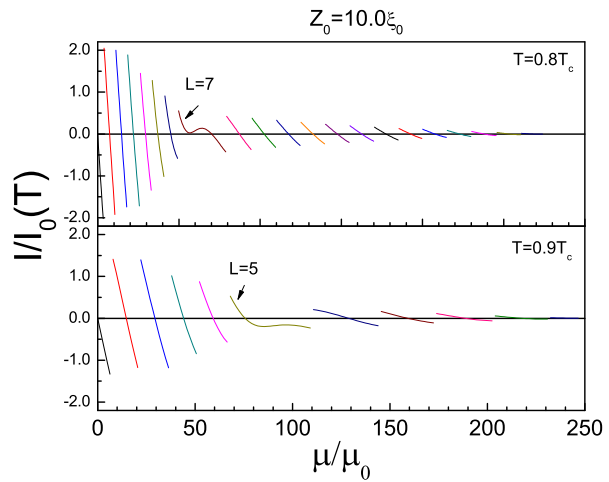}
\caption{(Color online) %
The current circulating in the tall cylinder (height
$Z_0=10.0\xi_0$) is shown here for the temperatures $T=0.8T_c$ and
$T=0.9T_c$. Current $L$ lines ranging are shown versus the magnetic
moment $\mu$ of the persistent current loop for the $L$ values of
Figs.(\ref{free_z10_t08}) and (\ref{free_z10_t09}). The single and
double well regimes reflect in current lines with different slopes
and the two regimes are separated by a transition line, $L=7$ for
$T=0.8T_c$, and $L=5$ for $T=0.9T_c$, respectively.
}\label{curr_z10_t08t09}
\end{figure}

\begin{figure}[b]
\includegraphics[width=1.0\linewidth]{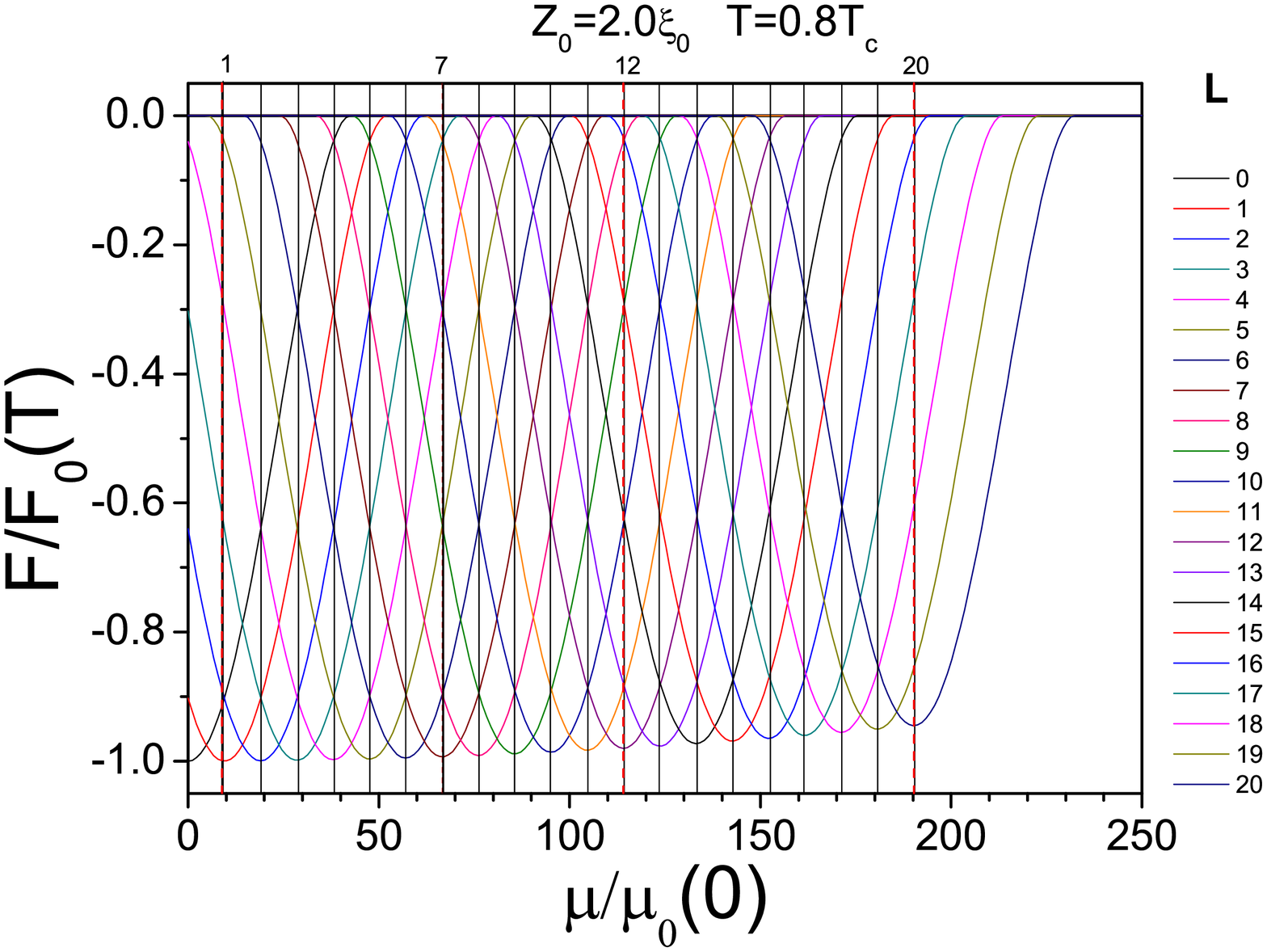}
\caption{(Color online) %
Vertical lines associated to the vanishing of the current are
plotted together with the free energy for the short cylinder (height
$Z_0=2.0\xi_0$) at the temperature $T=0.8T_c$. $L$ states ranging
from $0$ to $20$ are considered here. The $\mu$ values  equal to
$9.62$, $66.64$, $114.04$ and $190.29$, labeled as (1), (7), (12)
and (20), respectively, are depicted as red dashed vertical lines
and are selected for further analysis in
Fig.(\ref{vpj_z2_t08}).}\label{curr0_z2_t08}
\end{figure}

\begin{figure}[b]
\includegraphics[width=1.0\linewidth]{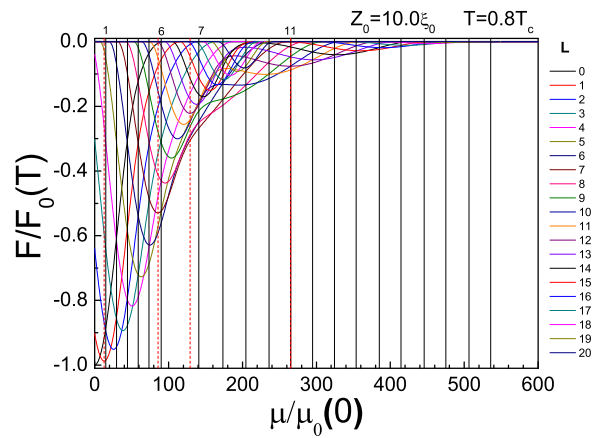}
\caption{(Color online) %
Vertical lines associated to the vanishing of the current are
plotted together with the free energy for the tall cylinder (height
$Z_0=10.0\xi_0$) at the temperature $T=0.8T_c$. $L$ states ranging
from $0$ to $20$ are considered here. The $\mu$ values equal to
$12.75$, $85.64$, $129.02$ and $264.97$, labeled as (1), (6), (7)
and (11), respectively, are depicted as red dashed vertical lines
and are selected for further analysis in
Fig.(\ref{vpj_z10_t08}).}\label{curr0_z10_t08}
\end{figure}

\begin{figure*}[!t]
\begin{center}
\includegraphics[width=0.8\linewidth]{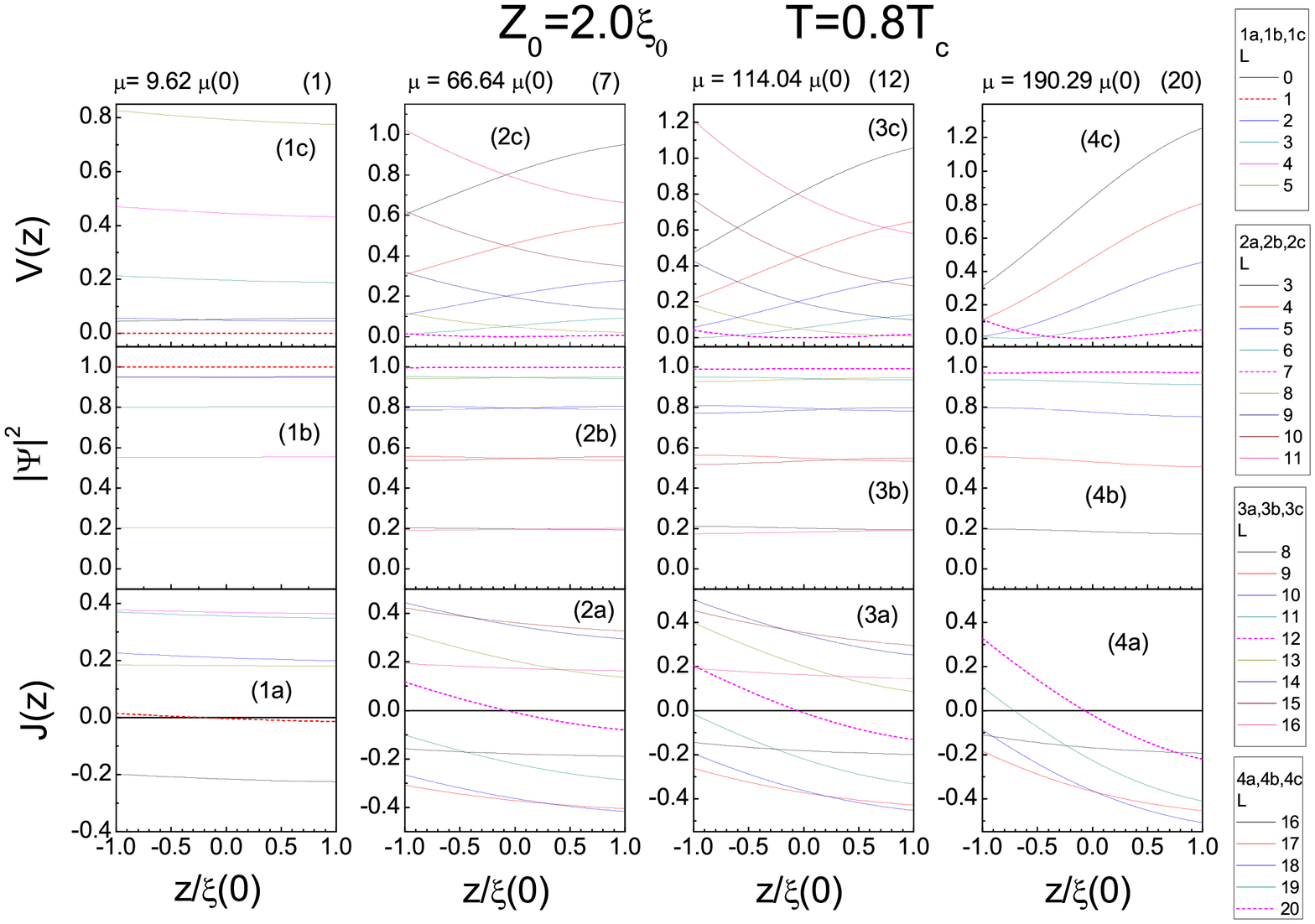}
\caption{(Color online) %
The current density $J$, the pair density $|\psi|^2$ and the
potential $V$ of the linear equation (Eq.(\ref{schro1})),  along the
cylinder height are shown here for the short cylinder (height
$Z_0=2.0\xi_0$) at the temperature of $T=0.8T_c$. The $\mu$ values,
selected in Fig.(\ref{curr0_z2_t08}), and  equal to $9.62$, $66.64$,
$114.04$ and $190.29$, labeled as (1), (7), (12) and (20), are
considered for this purpose. The persistent current loop is located
at position $1.5\xi_0$. The $L$ lines and their corresponding
figures are also listed here.}\label{vpj_z2_t08}
\end{center}
\end{figure*}

\begin{figure*}[!t]
\begin{center}
\includegraphics[width=0.8\linewidth]{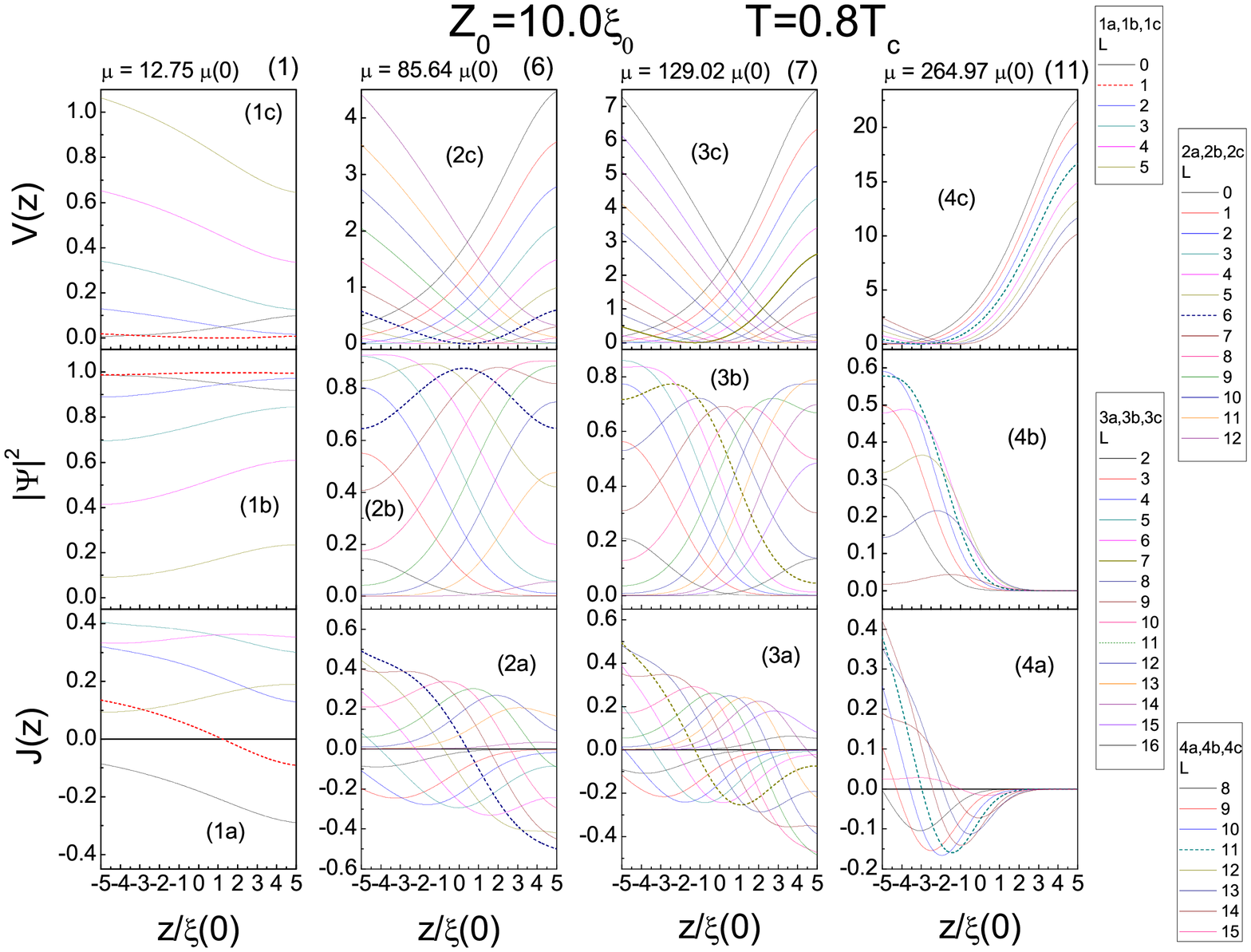}
\caption{(Color online) %
The current density $J$, the pair density $|\psi|^2$ and the
potential $V$ of the linear equation (Eq.(\ref{schro1})),  along the
cylinder height are shown here for the tall cylinder (height
$Z_0=10.0\xi_0$) at the temperature of $T=0.8T_c$. The $\mu$ values,
selected in Fig.(\ref{curr0_z10_t08}), and equal to $12.75$,
$85.64$, $129.02$ and $264.97$, labeled as (1), (6), (7) and (11),
are considered for this purpose. The persistent current loop is
located at position $5.5\xi_0$. The $L$ lines and their
corresponding figures are also listed here. }\label{vpj_z10_t08}
\end{center}
\end{figure*}

To describe the present system two other lengths must be considered,
namely, the distance from the current loop to the center of the
thin-walled cylinder, $z_1$ and the height of the thin-walled
cylinder, $Z_0=2z_0$, $z_0$ being the distance from the edge to its
center. Throughout this paper the distance between the current loop
and the top edge of the thin-walled cylinder is kept fixed and equal
to half coherence length, thus $z_1=0.5\xi_0+z_0$. Our analysis is
restricted to the thin-walled cylinder, which means that its width
is smaller than the coherence length, $w<\xi_0$ and therefore only
two coordinates are needed to determine a point in the cylinder,
namely, $(\rho=R,\varphi,|z| \le z_0 )$ in cylindrical coordinates
with the origin at the center of the thin-walled cylinder.  The
order parameter inside the thin-walled cylinder is expressed by its
phase $\phi$ and amplitude $f$, such that $\Psi=f(z)
\exp{\big(\imath\phi(\varphi,z)\big)}$. The well known condition of
a single value order parameter implies that
$\Psi(\varphi+2\pi)=\Psi(\varphi)$ and this means that the order
parameter must be the same along the $z$-axis: $\phi=L\varphi$,
where $L$ is an integer to be associated with the angular momentum
of the vortex state trapped in the thin-walled cylinder. Therefore
the order parameter is assumed to have the general form
$\Psi(z,\varphi)=f(z) \exp{\big(\imath L \varphi \big)}$. The
dipolar magnetic field falls with the inverse of the cube of the
distance to the current loop, thus being strong in the top and weak
at the bottom and this will make the amplitude strongly $z$
dependent. Although the current in the external loop always makes
the thin-walled cylinder feel a dipolar magnetic field, its
inhomogeneity is only noticeable in case the thin-walled cylinder is
sufficiently tall. For this reason we numerically study here two
examples of thin-walled cylinders here, a short ($Z_0=2.0\xi_0$) and
a tall ($Z_0=10.0\xi_0$) one. The former basically yields standard
LP oscillations and the latter comprises novel phenomena because the
top and the bottom edges of the thin-walled cylinder are pierced by
very distinct magnetic fluxes.

We solve the nonlinear GL equation for the thin-walled cylinder
through the finite element method (COMSOL and MATLAB softwares) and
check that our numerical procedure is correct in the short cylinder
limit, since in this case, we also solve the theory analytically.
This gives confidence that our numerical results for the thin-walled
cylinder are reliable.

The paper is organized as follows. In section~\ref{theoback} we
review the basic Ginzburh-Landau formalism applied to the
thin-walled cylinder in present of a persistent current loop. In
section~\ref{ring} we explore the limit that the amplitude $f$ can
be considered as constatn along the height to obtain important
theorems relating the zero current to the minimum of the free
energy. In section~\ref{numerical} we numerically study the short
and tall cylinders and find that the latter displays two distinct
regimes according to the the value of the persistent current.
Finally in section~\ref{conclusion} we reach some conclusion about
the results obtained here.

\section{Theoretical Background} \label{theoback}
The Ginzburg-Landau theory describes the superconductor near to its
critical temperature $T_{c}$. Because the London penetration length
is taken to be much larger than the hollow cylinder height, the
shielding of the magnetic field is safely ignored in our
calculations. Hence the free energy density  is simply that of the
sum of terms involving the order parameter:
\begin{eqnarray}
\mathcal{F}&=&\int_\Omega d\upsilon \Big[
\frac{\hbar^2}{2m}|\Big(\vec \nabla-\frac{2\pi
i}{\Phi_{0}}\vec{A}\Big)\Psi|^2+  \alpha(T)|\Psi|^2
+ \frac{\beta}{2}|\Psi|^4\Big] \nonumber \\\label{free0}
\end{eqnarray}
where $\alpha(T) \equiv \alpha_0\Big(\frac{T}{T_c}-1\Big)$,
therefore $\alpha(T) \le 0$ to sustain the superconducting state,
and $\beta$ is a constant. The volume of the thin-walled cylinder is
$\Omega\equiv 2\pi R w Z_{0}$ and the dimensionless volume element
is $d\upsilon\equiv d^3x/\Omega$. From the variational principle one
obtains the Ginzburg-Landau equation that determines the order
parameter which minimizes this free energy:
\begin{eqnarray}
-\frac{\hbar^2}{2m}\Big(\vec \nabla-\frac{2\pi
i}{\Phi_{0}}\vec{A}\Big)^2\Psi=- \alpha(T)\Psi + \beta |\Psi|^2\Psi
\label{gleq}
\end{eqnarray}
The persistent current loop has its center located at coordinates
$(0,0,z_1)$, its magnetic moment  is $\mu=I_{loop}\pi a^2$,
$I_{loop}$ being the circulating current. The vector potential
describing its field in space is given by
$\vec{A}(\rho,z)=A_{\varphi}(\rho,z)\hat \varphi$~\cite{selvaggi05},
\begin{eqnarray}
&&A_{\varphi}(\rho,z) \equiv \frac{\mu\rho}{[\rho^2+a^2+(z-z_1)^2]^{3/2}}\cdot  \nonumber \\
&&\cdot \sum_{n=0}^{\infty}\frac{(4n+1)!!}{2^{2n}n!(n+1)!}\Big(\frac{a\rho}{\rho^2+a^2+(z-z_1)^2}\Big)^{2n}= \nonumber \\
&&= \frac{\mu\rho}{[\rho^2+a^2+(z-z_1)^2]^{3/2}}\Big\{1+\frac{15}{8}q^2+\frac{315}{64}q^4+\cdots\Big\}, \nonumber  \\
&&q = \frac{a\rho}{\rho^2+a^2+(z-z_1)^2}.
\end{eqnarray}
The thin-walled condition restricts the radial coordinate to
$\rho=R$, and so, it suffices to use the expression,
\begin{eqnarray}
&&A_{\varphi}(R,z)= \frac{\mu}{R^2}\eta(z), \quad \mbox{where}\label{aphi}\\
&&\eta(z) \equiv
\frac{1}{\big[1+\big(\frac{a}{R}\big)^2+\big(\frac{z-z_1}{R}\big)^2\big]^{3/2}}\big(
1+1.8750 p^2+\nonumber \\
&&+4.9219 p^4+14.6631 p^6+46.7386 p^8+ 155.4058 p^{10}+ \nonumber \\
&&+ 531.8949 p^{12}+ 1.8593\,10^{3} p^{14}+6.6042\,10^{3}p^{16}+ \nonumber \\
&&+2.3757\,10^{4} p^{18}+8.6335\,10^{4} p^{20}+\cdots\big)
\label{eta} \\
&& p = \frac{\frac{a}{R}}{1 + \big(\frac{a}{R}\big)^2 +
\big(\frac{z-z_1}{R}\big)^2}.\label{p}
\end{eqnarray}
A term $c_{2n}\,p^{2n}$ of this series expansion has the coefficient
$c_{2n}$ growing with $n$, as shown above, that must be overcome by $p^{2n}$ in order to
achieve convergency. For instance, in the present case of interest,
$R=10.0\xi_0$, $a=5.0\xi_0$, and we take the top edge of the
cylinder, $z-z_1=-0.5\xi_0$ to obtain that $p=0.3992$.
Precision of the order $10^{-3}$ is only
achieve near the $n=10$ term when $c_{20}=8.6335\,10^{4}$ times $p^{20}=1.05\;10^{-8}$
gives $c_{20}\,p^{20}=9.0651\,10^{-4}$.

It is worthwhile to compute the
magnetic flux that pierces a ring resulting from the intersection of
the thin-walled cylinder with a plane $z$:
\begin{eqnarray}
\Phi(\mu,z) &\equiv& \oint_{z} d\vec l \cdot \vec A(R,z)= \nonumber
\\
&=&2\pi R A_{\varphi}(R,z)=\frac{2\pi\mu}{R}\eta(z)
\label{magflux}\\
\frac{\Phi(\mu,z)}{\Phi_0}&=&\frac{\mu}{\mu_0}\frac{\xi_0}{R}\eta(z)\label{magratio}
\end{eqnarray}
Eq.(\ref{magratio}) shows that the ratio between the magnetic
moments, $\mu/\mu_0$, together with  the geometrical parameters,
defines the ratio between the $z$ plane magnetic flux and the
fundamental flux. In order to gain insight into this series
expansion, let us analyze it in the case treated here, namely, the
ratio between the current loop and the thin-walled cylinder radii is
equal to one half, $a/R=0.5$. Suppose a point at the surface of the
thin-walled cylinder near to its top, such that $(z-z_1)/R\approx
0$. In this case we have that $p^2\approx 0.16$ and $
1+\frac{15}{8}p^2+\frac{315}{64}p^4+\cdots\approx
1+0.30+0.12+\cdots$. We learn that the series terms contribute in
this example, although they do not add any significant qualitative
change if instead we had considered the point like dipole limit $a
\rightarrow 0 $. Thus in this limit all the corrections in powers of
$(a/R)^2$ do not exist and the field is that of a point like
magnetic dipole, $ \vec \mu = \mu \hat z$ given by $\vec {A}=
\vec{\mu} \times \vec{r}/{\vert \vec r \vert}^3$ where $\vec r = R
\hat \rho + (z-z_1)\hat z$ falls in the surface of the thin-walled
cylinder.

We write the order parameter in terms of its amplitude $f$ and phase
$\phi=L\varphi$ to obtain the following expression for the kinetic
energy in terms of the geometry, the angular momentum $L$ and the
vector potential $A_{\varphi}$:
\begin{eqnarray}
G\equiv  \frac{\hbar^2}{2m}\int_\Omega d\upsilon
\Big\{\Big(\frac{\partial f}{\partial z} \Big)^2 +\frac{1}{R^2}
\Big[L-\frac{\Phi(\mu,z)}{\Phi_0}\Big]^2 f^2 \Big\}
\end{eqnarray}
Then the free energy density becomes,
\begin{eqnarray}
\mathcal{F}&=& G +\int_\Omega d\upsilon \Big[-\alpha(T) f^2 + \frac{\beta}{2} f^4\Big]
\end{eqnarray}
We also write the supercurrent density, $\vec{J}\equiv
\frac{q\hbar}{m} \Im \Big[ \Psi^*\Big(\vec \nabla-\frac{2\pi
i}{\Phi_{0}}\vec{A}\Big)\Psi \Big]$, in terms of the amplitude and
the phase decomposition using the cylindrical symmetry:
\begin{eqnarray}
\vec{J}= \frac{q\hbar}{mR}
[f(z)]^2\Big[L-\frac{\Phi_0(\mu,z)}{\Phi_0}\Big]\hat{\varphi}.
\label{currdens0}
\end{eqnarray}
There has been many proposed methods to calculate the current of the
LP effect~\cite{oppen92,conduit11}. In our numerical calculations
the current circulating in the thin-walled cylinder is simply
obtained as an integration of the current density, $\vec{J}$, along
the cross-section of the cylinder. According to the thin-walled
condition $w \sim \xi_0$, and so,
\begin{eqnarray}
I &=& w\int_{-z0}^{z0} dz \vec{J}\cdot \hat{\varphi} = \nonumber \\
 &=& I_0\int_{-z0}^{z0}
\frac{dz}{Z_0}\,\big[\frac{f(z)}{\sqrt{\alpha_0/\beta}}\big]^2
\Big[L-\frac{\Phi(\mu,z)}{\Phi_0}\Big], \nonumber \\ \label{curr0}
\end{eqnarray}
in terms of the current parameter $I_0 \equiv \big( q\hbar Z_0 w
\alpha_0 \big)/\big(m \beta R \big)$. The LP oscillations occur near
to the critical temperature, where the order parameter is weak, and
so, besides the non-linear, the linear version of Eq.(\ref{gleq}),
with the cubic term neglected, can be used. Then there is a
Schr\"odinger like equation for the amplitude $f(z)$, once the order
parameter is taken as $\Psi(z,\varphi)=f(z) \exp{\big(\imath L
\varphi \big)}$:
\begin{eqnarray}
\big[-\frac{\hbar^2}{2m}\frac{d^2}{dz^2}
+V(z)\big]f(z)=-\alpha(T)f(z), \label{linearf}
\end{eqnarray}
where the potential is given by
\begin{eqnarray}
V(z) = \frac{\hbar^2}{2mR^2}
\big[L-\frac{\Phi(\mu,z)}{\Phi_0}\big]^2, \label{pot1}
\end{eqnarray}
and the boundary condition at the edges of the thin-walled cylinder
are $df(z)/dz|_{(z=\pm z_0)}=0$, which is automatically satisfied in
the limit of a constant $f$. Let us consider the ill-defined limit,
but still instructive, of $f$ $z$ independent, such that
$d^2f(z)/dz^2\approx 0$. Then the above equation defines a condition
under the simple assumption that there is a non-zero $f$:
$V(z)+\alpha(T)=0$ or equally, $T/T_c = 1-\big(\xi_0/R\big)^2
\big[L-\Phi(\mu,z)/\Phi_0\big]^2$. This defines the LP temperature,
which is a function of the flux that pierces the thin-walled
cylinder along the $z$ plane. However because $\Phi(\mu,z)$ is $z$
dependent, the found ratio $T/T_c$ is in contradiction with the
original argument. Indeed a $z$ independent amplitude $f$ is only
possible for a very short thin-walled cylinder, and this entitles us
to take $z=0$, which means the medium plane. We choose to express
the above ratio $T/T_c$ in terms of the $T=0$ coherence length,
$\xi_0\equiv\sqrt{\hbar^2/2m\alpha_0}$. The dimensionless ratio
$\Phi(\mu,z)/\Phi_0$ can be written in terms of the dimensionless
length ratios, $R/\xi_0$, $a/R$, and $(z-z_1)/R$, and the
dimensionless magnetic moment ratio, $\mu/\mu_0$ where $\mu_0$ has
been previously defined. This naive approach to the LP oscillations
is improved in the next sections by treating the $z$ dependence of
this problem. The thin-walled cylinder in presence of the
inhomogeneous field produced by the persistent current loop field
makes the amplitude dependent on its height position: $f(z)$. In
this case we also solve the above Schr\"oedinger equation in order
to determine $f(z)$.

\section{The ring limit} \label{ring}
A thin-walled cylinder such the order parameter amplitude $f$ is
constant along the $z$ direction, and therefore $\partial f/
\partial z=0$, we define as a ring. The validity of the ring treatment
restricts the height to a few coherence length units, which means
that the ratio $z_0/\xi_0$ cannot be very large. In the ring limit
all that is necessary to determine the order parameter and the free
energy is to compute the integrals,
\begin{eqnarray}
g(\mu) &\equiv&  \int_{-z0}^{z0} \frac{dz}{Z_0} \Big[L-\frac{\Phi(\mu,z)}{\Phi_0}\Big]^2, \quad \mbox{and}, \label{g}\\
g'(\mu) &\equiv&  \int_{-z0}^{z0} \frac{dz}{Z_0}
\Big[L-\frac{\Phi(\mu,z)}{\Phi_0}\Big]. \label{gp}
\end{eqnarray}
The kinetic energy energy  becomes $G =
\alpha_0\big(\frac{\xi_0}{R}\big)^2 g f^2$ and the current
$I=\frac{q\hbar}{m}\frac{wZ_0}{R} f^2 g'$. Then, we have for the
free energy,
\begin{eqnarray}
\mathcal{F}=[\alpha_0\big(\frac{\xi_0}{R}\big)^2g+\alpha(T)] f^2+
\frac{\beta}{2} f^4
\end{eqnarray}
whose minimization determines the order parameter and its minimum:
\begin{eqnarray}
f^{2}&=&-\frac{1}{\beta}\big[\alpha_0\big(\frac{\xi_0}{R}\big)^2g+\alpha(T)\big] \nonumber\\
\mathcal{F}
&=&-\frac{1}{\beta}\big[\alpha_0\big(\frac{\xi_0}{R}\big)^2g+\alpha(T)\big]^2
\label{f0}
\end{eqnarray}
Notice that since $ g > 0$ and the above equation determines that
$\alpha_0\big(\frac{\xi_0}{R}\big)^2 g+\alpha(T) \le 0$ because
$f_{0}^2>0$ then there is a temperature bound given by $\alpha(T)\le
-\alpha_0\big(\frac{\xi_0}{R}\big)^2 g$. The upper bond defines the
LP temperature, which corresponds to the vanishing of the order
parameter, such that the superconductor reaches the normal state for
a temperature which depends on $g$ and the $\alpha_0$ parameter,
according to the above equation. Therefore the LP oscillations
obtained from the non-linear theory in case of $f$ constant are
expressed in terms of the integrals $g$ and $g'$. The corresponding
free energy, current  and critical temperature are defined as
follows:
\begin{eqnarray}
\mathcal{F}(\mu,T)&=& - \mathcal{F}_0\Big[
\frac{T-T_{c}(\mu)}{T_{c}}\Big]^2 \label{free}\\
\frac{T_c(\mu)}{T_c}&=&1-\big(\frac{\xi_0}{R}\big)^2 g(\mu) \label{temp}\\
I(\mu,T)&=&-I_0\Big[ \frac{T-T_{c}(\mu)}{T_{c}}\Big]g'(\mu)
\label{curr}
\end{eqnarray}
The integrals $g(\mu)$ and $g'(\mu)$ can be expressed in terms of
the following defined integrals:
\begin{eqnarray}
\langle L \rangle &\equiv&  \int_{-z0}^{z0} \frac{dz}{Z_0}
\frac{\Phi(\mu,z)}{\Phi_0}, \quad \mbox{and}, \\
\langle L^2 \rangle &\equiv&  \int_{-z0}^{z0} \frac{dz}{Z_0}
\Big(\frac{\Phi(\mu,z)}{\Phi_0}\Big)^2,
\end{eqnarray}
such that, the two integrals can be written as,
\begin{eqnarray}
\langle L \rangle &=&  \frac{\mu}{\mu_0}I_1, \quad I_1 \equiv
\frac{\xi_0}{R}\int_{-z0}^{z0} \frac{dz}{Z_0}\eta(z),\quad \mbox{and},\\
\langle L^2 \rangle &=&  \big(\frac{\mu}{\mu_0}\big)^2 I_2, \quad
I_2 \equiv \big(\frac{\xi_0}{R}\big)^2\int_{-z0}^{z0}
\frac{dz}{Z_0}\big[\eta(z)\big]^2
\end{eqnarray}
By taking that $\big(\frac{\xi_0}{R}\big)^2\int_{-z0}^{z0}
\frac{dz}{Z_0}\big[\eta(z)-I_1\big]^2 \ge 0$, we obtain that
\begin{eqnarray}
I_2-I_1^2 \ge 0.\label{ineq}
\end{eqnarray}
Next we show three important features of the LP oscillations with
respect to the magnetic moment $\mu$ that can be expressed in terms
of the above functions $g(\mu)$ and $g'(\mu)$, written in the
following way:
\begin{eqnarray}
g(\mu) &=& L^2 - 2 L\frac{\mu}{\mu_0}I_1 + \big(\frac{\mu}{\mu_0}\big)^2 I_2, \quad \mbox{and}, \\
g'(\mu) &=& L - \frac{\mu}{\mu_0}I_1.
\end{eqnarray}
These three three features are valid to all temperatures, and for
this reason assume that $T \ne T_c(\mu)$. They are associated to
three special moments, $\mu_f$, $\mu_c$ and $\mu_t$, defined as
such:
\begin{enumerate}
\item[1)] {\it The minimum of the free energy}.
According to Eq.(\ref{free}), $\partial \mathcal{F}/\partial\mu=0$
is the same condition as the minimum with respect to the critical
temperature, $\partial T_c(\mu)/\partial\mu=0$, which boils down to
$\partial g(\mu)/\partial\mu=0$. This defines the magnetic moment
$\mu_f$.
\begin{eqnarray}
\frac{\mu_f}{\mu_0}= L\frac{I_1}{I_2} \label{muf}
\end{eqnarray}
\item[2)] {\it The vanishing of the current}.
According to Eq.(\ref{curr}) the condition $I(\mu)=0$ happens for
$g'(\mu)=0$. This defines the magnetic moment $\mu_c$:
\begin{eqnarray}
\frac{\mu_c}{\mu_0}= L\frac{1}{I_1} \label{muc}
\end{eqnarray}
\item[3)]{\it The highest LP temperature}.
According to Eq.(\ref{temp}) this occurs when the LP temperature
becomes the critical one. However the condition $Tc(\mu)/T_c=1$ is
the same as $g(\mu)=0$. This defines the magnetic moment $\mu_t$:
\begin{eqnarray}
\frac{\mu_t}{\mu_0}= L\frac{I_1}{I_2} \Big(1\pm \imath
\sqrt{\frac{I_2}{I_1^2}-1} \Big) \label{mut}
\end{eqnarray}
\end{enumerate}
Then we prove that {\it the LP temperature never reaches the
critical temperature $T_c$}, unless for the zero height ring, where
$\Phi_0(\mu,z)=\Phi_0(\mu,0)$ ($I_2=I_1^2$). Notice that
\begin{eqnarray}
\frac{\mu_c}{\mu_f}= \frac{I_2}{I_1^2}\ge 1,
\end{eqnarray}
which means that {\it the current in the thin-walled cylinder always
vanishes after the free energy reaches its minimum ($\mu_c \ge
\mu_f$)} by increasing the current in the loop. Another important
aspect to consider is that the three special magnetic moments,
defined by Eqs.(\ref{muf}), (\ref{muc}) and (\ref{mut}), are
temperature independent. This means that {\it the distance between
two consecutive $L$ and $L+1$ magnetic moments, either free energy
minima or zeros of the current, are temperature independent and
equal to $I_1/I_2$ and $1/I_1$}, respectively. For a very short
ring, defined to have height comparable to the coherence length,
$Z_0\le \xi_0$, the integrals $I_1$ and $I_2$ are easily computed,
since it is enough to take the integrands at the center of the ring,
$z \approx 0$, to obtain that $I_1\approx \eta(0) $, and $I_2\approx
\eta(0)^2$. The three conditions, namely, the minimum of the free
energy, the vanishing of the current and the highest LP temperature
take place at the same point $\mu/\mu_0=L/I_1$ because $I_1^2 =
I_2$.  Inserting the integrals and using the magnetic flux
definition of Eq.(\ref{magflux}) in the center plane of the ring we
obtain for the free energy density, critical temperature and current
density the following expressions:
\begin{eqnarray}
\mathcal{F}(\mu,T)&=& - \mathcal{F}_0\Big[
\frac{T-T_{c}(\mu)}{T_{c}}\Big]^2\\
\frac{T_{c}(\mu)}{T_{c}}&\equiv&
1-\Big(\frac{\xi_0}{R}\Big)^2\Big[L-\frac{\Phi(\mu,0)}{\Phi_0}\Big]^2 \label{temp1}\\
I(\mu,T)&=&-I_0\Big[
\frac{T-T_{c}(\mu)}{T_{c}}\Big]\Big[L-\frac{\Phi(\mu,0)}{\Phi_0}\Big]\\
\end{eqnarray}
where we have defined the parameters, associated to the free energy,
$\mathcal{F}_0\equiv \alpha_0^2/\beta $, and to the current, $I_0$,
previously defined. Notice that the above LP temperature was
previously obtained using the linear approach. The above case is the
standard LP effect. valid for a constant applied external field and
arbitrary height.

To show how the height of the cylinder affects the LP oscillations
we discuss below the approximation that $f$ is still a constant, but
the integrals $g$ and $g'$ must be integrated along the $z$ axis. We
get analytical insight into this problem by taking that the current
loop is very small as compared to the thin-walled cylinder, $a/R
<<1$, such that just the first term of the series expansion given in
Eq.(\ref{aphi}) is enough to describe the external magnetic field:
\begin{eqnarray}
\eta(z) = \frac{1}{[1+\big(\frac{z-z_1}{R}\big)^2]^{3/2}}.
\end{eqnarray}
This vector potential is introduced into Eqs.(\ref{g}) and (\ref{gp})
to obtain that,
\begin{eqnarray}
I_1= \frac{\xi_0}{Z_0}
\big\{\frac{\big(\frac{z_0-z_1}{R}\big)}{\sqrt{1+\big(\frac{z_0-z_1}{R}\big)^2}}
+\frac{\big(\frac{z_0+z_1}{R}\big)}{\sqrt{1+\big(\frac{z_0+z_1}{R}\big)^2}}\big\}
\end{eqnarray}
\begin{eqnarray}
I_2 &=& \frac{\xi_0^2}{Z_0 R} \big\{\frac{1}{4}\big[ \frac{\big(\frac{z_0-z_1}{R}\big)}{\big(1+\big(\frac{z_0-z_1}{R}\big)^2\big)^2}+ \frac{\big(\frac{z_0+z_1}{R}\big)}{\big(1+\big(\frac{z_0+z_1}{R}\big)^2\big)^2}\big]+ \nonumber\\
&+&  \frac{3}{8}\big[ \frac{\big(\frac{z_0-z_1}{R}\big)}{1+\big(\frac{z_0-z_1}{R}\big)^2}+ \frac{\big(\frac{z_0+z_1}{R}\big)}{1+\big(\frac{z_0+z_1}{R}\big)^2}\big]+ \nonumber \\
&+&  \frac{3}{8}\big[ \arctan{\big(\frac{z_0-z_1}{R}\big)} + \arctan{\big(\frac{z_0+z_1}{R}\big)} \big] \big\}\nonumber \\
\end{eqnarray}
These integrals are expanded in a Taylor series in powers of
$(z_0/R)^2$, which is a small parameter, since the approximation of
$f$ constant is a good one only for a short cylinder. We obtain for
the integrals $I_1$ and $I_2$ the following values by keeping
contributions of lowest order $(z_0/R)^2$ and neglecting higher
ones.
\begin{eqnarray}
I_1= \frac{\xi_0}{R} \frac{1}{\big[1+\big(\frac{z_1}{R}\big)^2 \big]^{3/2} }\big\{1 -\frac{1}{2}\frac{1-4\big(\frac{z_1}{R}\big)^2}{\big[1+\big(\frac{z_1}{R}\big)^2 \big]^{2} } \big(\frac{z_0}{R}\big)^2\big\} \nonumber \\
I_2= \big(\frac{\xi_0}{R}\big)^2 \frac{1}{\big[1+\big(\frac{z_1}{R}\big)^2 \big]^{3} }\big\{1 -\frac{1-7\big(\frac{z_1}{R}\big)^2}{\big[1+\big(\frac{z_1}{R}\big)^2 \big]^{2} } \big(\frac{z_0}{R}\big)^2\big\} \nonumber \\
\end{eqnarray}
We notice that the above equations do satisfy the inequality of
Eq.(\ref{ineq}),
\begin{eqnarray}
I_2-I_1^2= 3
\frac{\big(\frac{\xi_0}{R}\big)^2\big(\frac{z_0}{R}\big)^2\big(\frac{z_1}{R}\big)^2}{\big[1+\big(\frac{z_1}{R}\big)^2
\big]^{5}}
\end{eqnarray}
As an example let us consider the short cylinder with radius
$R=10.0\xi_0$, half-height $z_0=1.0\xi_0$ and the current loop just
above its top, $z_1=1.5\xi_0$. Then $I_2-I_1^2\sim
\big(\frac{\xi_0}{R}\big)^2\big(\frac{z_0}{R}\big)^2\big(\frac{z_1}{R}\big)^2\sim
10^{-6}$. The distance between two consecutive free energy minima is
10.2456$\mu_0$ and between two consecutive current zeros is
10.3507$\mu_0$.
\section{Numerical Analysis}
\label{numerical}
We obtain numerical solutions of the non-linear GL equation below,
\begin{eqnarray}
\big[-\frac{\hbar^2}{2m}\frac{d^2}{dz^2}
+V(z)\big]f(z)=-\alpha(T)f(z)+\beta f(z)^3.\label{nonlf}
\end{eqnarray}
by solving it through the finite element method (COMSOL and MATLAB
softwares). Then we obtain the free energy, the current density and
the current from Eqs.(\ref{free0}), (\ref{currdens0}), and
(\ref{curr0}), respectively, by input of $f(z)$, obtained from
Eq.(\ref{nonlf}).

To obtain the numerical solutions of Eq.(\ref{nonlf}) we cast it in
dimensionless quantities, represented by a bar on top of the corresponding
variable: $\bar f = \sqrt{\beta/\alpha(T)}f$, $\bar z = z/\xi(T)$,
where $\xi(T)=\xi_0/\sqrt{1-T/T_c}$, and $\bar V = V/\alpha(T)$.
The explicit temperature dependence disappears from
Eq.(\ref{nonlf}) which becomes,
\begin{eqnarray}
\big[-\frac{d^2}{d{\bar z}^2} +\bar V(\bar z)\big]\bar f(\bar
z)=-\bar f(\bar z)+\bar f(\bar z)^3.\label{nonlf1}
\end{eqnarray}
In practical grounds this means that all the lengths are measured in
units of $\xi(T)$, the temperature dependent coherence length, and
so, ratios between two lengths become temperature independent. For
instance, the function $\eta$, defined by Eq.(\ref{eta}), is found
to be temperature independent, $\eta(\bar z)=\eta(z)$, since it only
depends on ratios. The dimensionless version of the potential,
defined in Eq.(\ref{schro1}), becomes temperature dependent and given
by,
\begin{eqnarray}
\bar V(\bar z)= \frac{1}{\bar R(T)^2} \big [ L- \bar \mu(T)
\frac{\eta(\bar z)}{\bar R(T)}\big]^2.\label{schro1}
\end{eqnarray}
where $\bar R(T) = R/\xi(T)$, $\bar \mu(T) = \mu/\mu_0(T)$ and
$\mu_0(T)=\Phi_0\xi(T)/2\pi$.
Notice that the product $\mu_0(T)\bar R(T) = \Phi_0 R/2\pi$ is temperature independent.
Like the radius, the height of the cylinder also shrinks to zero as $T$ approaches $T_c$:
$\bar z_0=z_0/\xi(T)$.

The free energy, the current density and the current, obtained from
Eqs.(\ref{free0}), (\ref{currdens0}), and (\ref{curr0}),
respectively, by input of $f(z)$, determined from Eq.(\ref{nonlf}),
are also cast into dimensionless units. The current is given by,
\begin{eqnarray}
\bar I =\int_{-\bar z_0}^{\bar z_0} \frac{d\bar
z}{\bar Z_0}\,\bar f(\bar z)^2 \big [ L- \bar \mu(T) \frac{\eta(\bar
z)}{\bar R(T)}\big]\label{curr1}
\end{eqnarray}
where $\bar I = I/I_0(T)$,  $I_0(T) \equiv \big( q\hbar \bar
Z_0 w \alpha(T) \big)/\big(m \beta \bar R \big)$.
The free energy in reduced units becomes,
\begin{eqnarray}
\bar F = \int_{-\bar z_0}^{\bar z_0} \frac{d\bar z}{\bar Z_0}\, \big\{
\big[ \frac{d\bar f(\bar z)}{d \bar z}\big]^2 + \big [ \bar V(\bar
z)-1 \big ] \bar f(\bar z)^2 +\frac{1}{2}\bar f(\bar z)^4 \big\}
\nonumber \\ \label{free1}
\end{eqnarray}
where $\bar F = F/F_0(T)$, $F_0(T)\equiv\alpha(T)^2/2\beta$.

We concentrate our numerical study in two thin walled cylinders,
both with the same radius $R=10.0\xi_0$ in presence of the same
current loop ring with radius $a=5.0\xi_0$, set at $0.5\xi_0$ above
the top edge. The two cylinders only differ by the height, taken to
be $Z_0=2.0\xi_0$ and $Z_0=10.0\xi_0$, such that $z_1=1.5\xi_0$ and
$5.5\xi_0$, respectively. We apply the same numerical method to
both, but the short cylinder is expected to have the amplitude $f$
nearly $z$ independent and be approximately described by the
analytical ring limit previously considered. From the other side the
tall cylinder, $Z_0=10.0\xi_0$, can only be described by the full
numerical treatment since the amplitude varies along the cylinder's
height, $f(z)$. All the free energy and the current plots are in
terms of the $T=0$ magnetic moment $\mu/\mu_0(0)$, in order to have
a temperature independent scale. Notice that
$\mu_0(T)=\Phi_0\xi(T)/2\pi$ and we are taking its zero temperature
value as our magnetic moment scale.

Figs.~(\ref{lin_z2}) and (\ref{lin_z10}) show the LP temperature
versus magnetic moment temperature diagram derived from the linear
theory. The ratio $T_c(\mu)/T_c$ is calculated  from the lowest
eigenvalue of Eq.(\ref{nonlf1}) and (\ref{pot1}) for given $L$ and
$\mu/\mu_0(0)$ values. These $L$ lines set the border line that
separates the superconducting to the normal state in the diagram.
The two cylinders display very distinct $T_c(\mu)/T_c$ versus
$\mu/\mu_0(0)$ curves, as shown in Figs.~(\ref{lin_z2}) and
(\ref{lin_z10}).

Fig.~(\ref{lin_z2}) covers the $0$ to $200$ $\mu/\mu_0(0)$ range,
which contains the first $20$ $L$ lines of the $Z_0=2\xi_0$
cylinder. Notice the parabolic shape of the $L$ lines and their
nearly homogeneous spatial distribution. The distance between two
consecutive maxima, of $L$ and $L+1$ lines, is described by the
formula $\Delta \mu/\mu_0(0) \approx 9.5$. Notice that for
increasing magnetic moment the ratio $T_c(\mu)/T_c$ becomes lower
than one, indicating that the height of the cylinder matters in this
case, and our theoretical considerations developed in the ring limit
applies to this situation and is determined by Eq.(\ref{temp}) and
not by Eq.(\ref{temp1}). The ratio $T_c(\mu)/T_c$ reaches one only
in case there is only a single magnetic flux piercing the ring, that
is, in case of the nearly zero height cylinder or of the constant
magnetic field applied to cylinder of a arbitrary height.

Fig.~\ref{lin_z10} shows a much richer structure for the $L$ lines
of the $R=10\xi_0$ cylinder, due to the inhomogeneity of the
magnetic field that results in distinct magnetic fluxes piercing
this cylinder, each for a different $z$ plane intersect. Notice the
presence of an inner and an external (true) border lines. To
understand the $R=10\xi_0$ diagram, we selected a few $L$ lines,
plotted as thick lines for best visualization purposes: $L=5$, $12$,
$20$ and $41$. The $L=12$ line is a key case because of its double
peak structure with a local minimum between them, a common feature
to many lines. The first peak is narrow and tall, while the second
peak is short and broad and takes place at a larger magnetic moment
than the first peak. The tips of the narrow and broad peaks of the
$L=20$ line still fall inside the plotted temperature range, $0.8$
to $1.0$ $T_c(\mu)/T_c$. The $L=5$ thick line shows the onset of the
second peak, not present in the lower $L$ lines, which only have a
single peak. The $L=5$ and $6$ can be regarded as transition lines
to the onset of the second peak and $L=7$ is the critical line
because the second peak is clearly established for $L=8$ and beyond.
$L=41$ is the highest line in this diagram, seen around
$\mu/\mu_0(0)=400$ and setting the end of an inner transition line
inside the superconducting region. The first (narrow) peak
contributes to the true border line only up to $L=4$ then to
submerge and form an inner border line for high $L$ values. Thus
this inner border line is the curve tangent to the maximum of the
first peaks. From its side, the second (broad) peak only contributes
to the true superconducting normal state border line, from the $L=5$
line up to $L=20$ line. For the sake of the argument we have
included the $L=5,6$ and $7$ lines in this set, although the second
peak is only in an embryonic level there. In conclusion the
superconducting normal state border line has less angular momentum
states for the tall cylinder, since $L=20$ is the maximum possible
value reached at $\mu/\mu_0=510$, whereas for the short cylinder,
$L=20$ is reached for $\mu/\mu_0=200$.

The full non-linear theory is used to obtain the next plots, which
means that Eq.(\ref{nonlf1}) is numerically solved by means of the
finite element method, and next the free energy and current
calculated from Eqs.(\ref{free1}) and (\ref{curr1}), respectively.
In the dimensionless treatment the temperature $T$ is introduced by
adjustment of the radius and of the height, given by $\bar R(T)$ and
$\bar Z_0(T)$, respectively.

Fig.~\ref{free_z2_t08t09} shows several $L$ free energy lines of the
$Z_0=2.0\xi_0$ cylinder for the two temperatures of $T/T_c=0.8$ and
$0.9$. The minimum of each $L$ line is also a free energy minimum.
These minima are equally spaced and two consecutive ones, associated
to $L$ and $L+1$ lines, are separated by $\Delta \mu/\mu_0(0)
\approx 9.5$ in agreement with our theoretical analysis that
predicts $\Delta \mu/\mu_0(0) = I_1/I_2$, according to
Eq.(\ref{muf}). This position is temperature independent and
Fig.~\ref{free_z2_t08t09} has a dashed straight line to indicate
that for the two temperatures the minimum is in the same magnetic
moment value.

Fig.~\ref{curr_z2_t08t09} shows the vanishing of the current of each
$L$ line for the two temperatures of $T/T_c=0.8$ and $0.9$. The
zeros are found to be equally spaced, $\Delta \mu/\mu_0(0) \approx
9.5$, also in agreement with our theoretical analysis that predicts
$\Delta \mu/\mu_0(0) = 1/I_1$, according to Eq.(\ref{muc}). The
dashed vertical straight line is a guide to the eye to show that the
current vanishes at the same magnetic moment independently of the
temperature.

Fig.~\ref{free_z10_t08} shows the free energy of the tall cylinder
($Z_0=10.0\xi_0$) for the temperature of $T/T_c=0.8$. Distinctly
from the short cylinder case, here the minimum of a $L$ line is not
a free energy minimum, unless approximately for the first few lines,
$L=0$ to $4$. The $L$ lines undergo a transition in shape from the
small to the large $L$ limits. To understand this transition we
selected the $L=3$ and $12$ lines, also shown in the inset, as they
are representative of each side of the transition. While the $L=3$
line has a single minimum, $L=12$ has double well shape with two
minima at $\mu/\mu_0(0) \approx 130$ and $270$, the first one being
the lowest one in free energy. Nevertheless the second well is more
important than the first one because it sets the true free energy
minimum at $\mu/\mu_0(0) \approx 300$, thus away from the $L$ line
secondary local minimum. The first one has other competing states
with lower energy, the lowest one being the $L=7$ state, which is
drawn as a thick line to help the visualization. The $L=7$ state is
a critical one as it sets the onset of the double well structure.
This double well structure is not a feature present in all lines,
the first ones simply do not have it. In conclusion the free energy
clearly shows two types, single and double well shaped, that
dominate the small and large magnetic moment regimes, respectively.
This behavior is also seen in the temperature $T/T_c=0.9$, as shown
in Fig.\ref{free_z10_t09}. There $L=5$ is the critical line, and we
have also plotted as thick lines, the $L=3$ and $L=8$, to show the
single to double well transition.


Fig.~\ref{curr_z10_t08t09} shows that the $Z_0=10.0\xi_0$ cylinder
displays novel features as compared to the standard LP problem. The
current of the transition $L$ line that separates the single (low
$L$) to the double well (high $L$) behavior displays unusual
behavior. Both limits are characterized by distinct slopes but apart
from this both single and double well regions are similar. as shown
for the , but  not in the transition ($L=7$) line. This is seen for
both temperatures, $T/T_c=0.8$ and $0.9$. The transition lines occur
for $L=7$ and $5$, respectively and displays a very unusual current
pattern near its zero.

The zeros of the current and the free energy minima coincide for the
standard LP, but the inhomogeneous external field sets them apart,
as shown here. For this reason we plot the free energy for a given
temperature ($T=0.8T_c$) together with straight vertical lines
associated to the zeros of the current. Fig.~\ref{curr0_z2_t08}
shows for the $Z_0=2.0\xi_0$ cylinder that these straight vertical
lines fall very close to the free energy minima in agreement with
our theoretical predictions derived in the ring limit. We have
selected a few magnetic moment values (red dashed vertical lines) in
this diagram for further analysis, given by $\mu/\mu_0=9.62$,
$66.64$, $114.04$ and $190.29$, whose free energy minima correspond
to the $L=$ $1$, $7$ $12$ and $20$ lines. Interestingly we find that
these vertical lines intercept the matching points between the $L-1$
and $L+1$ free energy lines up to numerical precision. The distance
between consecutive vertical lines is in agreement with the
theoretical prediction for $\Delta\mu_c / \mu_0 (0) \approx 9.5$. We
observe that this property seems to be general, namely, also valid
for the $Z_0=10.0\xi_0$ cylinder, as shown in
Fig.~\ref{curr0_z10_t08}. However for the tall cylinder the
situation is quite different as the vertical lines clearly form two
sets, a low and a high magnetic moment set with a gap between them,
located at the $L=7$ line. The two sets feature distinct separation
between consecutive vertical lines with very distinct behavior
regarding the free energy minima. The vertical lines of
Fig.~\ref{curr0_z10_t08} split into two sets according to their
separation. For low values of magnetic moment(until $L=5$) the
distance between two consecutive lines is $\Delta \mu / \mu_0 (0)
\approx 14.7$. Then, there is a transition region($L=6,7,8$), and
for $L>9$ the curves of the free energy are dominated by the second
well. In this region, the distance between the lines is given by
$\Delta \mu / \mu_0(0) \approx 30.1$. Comparison betwee tall and
short cylinders show that their low magnetic moment has similar
behavior, but not for the high magnetic moment regime because of the
double well shape of the $L$ lines for the tall cylinder. The zeros
of the current are far from the free energy minima and yet they
coincide with the crossing of the $L-1$ and $L+1$ free energy lines.
In this case we have also selected a few magnetic moment values (red
dashed lines) for further analysis: $\mu/\mu_0=12.75$, $85.64$,
$129.02$ and $264.97$, They are labeled $1$, $6$, $7$ and $11$
because they fall within the regimes where these $L$ lines provide
the free energy minima.

In the ring limit (very short cylinder) it is easy to verify that
the crossing point between any $L-1$ and $L+1$ free energy curves
happens at the magnetic moment value of the vanishing of the $L$
state current. In the ring limit the amplitude $f$ is constant and
the free energies $\mathcal{F}_{L-1}$ and $\mathcal{F}_{L+1}$ for
states $L-1$ and $L+1$, respectively, are given by,
\begin{eqnarray}
&&\mathcal{F}_{L-1}=-\frac{1}{\beta}[\alpha_0 (\frac{\xi_0}{R})^2 g_{L-1}+\alpha(T)]^2, \quad \mbox{and} \\
&&\mathcal{F}_{L+1}=-\frac{1}{\beta}[\alpha_0 (\frac{\xi_0}{R})^2,
g_{L+1}+\alpha(T)]^2
\end{eqnarray}
according to Eq.(\ref{f0}). Next take that
$\mathcal{F}_{L-1}=\mathcal{F}_{L+1}$ to obtain that
$g_{L-1}(\mu^{*})=g_{L+1}(\mu^{*})$, which gives that,
\begin{eqnarray}
&&\int_{-z0}^{z0}\frac{dz}{Z_0}\Big\{\Big[L-1-\frac{\Phi(\mu^{*},z)}{\Phi_0}\Big]^2\
- \nonumber \\
&& - \Big[L+1-\frac{\Phi(\mu^{*},z)}{\Phi_0}\Big]^2\Big\}=0,
\nonumber \\
\end{eqnarray}
that once simplified becomes,
\begin{eqnarray}
\int_{-z0}^{z0} \frac{dz}{Z_0}
\Big[L-\frac{\Phi(\mu^{*},z)}{\Phi_0}\Big]=0
\end{eqnarray}
Inserting the definition of the magnetic flux $\Phi(\mu^{*},z)$
(Eq.(\ref{magflux})), and expressing the result as a function of
integral $I_{1}$, previously defined, we obtain that,
\begin{eqnarray}
\frac{\mu^{*}}{\mu_0}=L\frac{1}{I_1}.
\end{eqnarray}
This shows that for constant amplitude $f$ it holds that
$\mu^{*}=\mu_c$, according to Eq.(\ref{muc}). We also stress that
our present numerical evidence shows that this property also holds
for the tall cylinder where the amplitude $f$ is no longer constant.


To gain further insight into the structure of the superconducting
state along the cylinder height, we plot in reduced units the
current density $J(z)$, the superconducting density $|\psi(z)|^2$,
and the potential $V(z)$ as a function of $z$. Notice that the
current loop, which is the source of the inhomogeneous magnetic
field, is located in the positive $z$ axis $0.5\xi_0$ above the
maximum height. Fig.~\ref{vpj_z2_t08} shows these plots for the
$Z_0=2.0\xi_0$ cylinder at the selected magnetic moment values
called $(1)$, $(7)$, $(12)$ and $(20)$. All the possible crossing
$L$ states with the vertical lines, as shown in
Fig.~\ref{curr0_z2_t08}, are in the $J(z)$, $|\psi(z)|^2$ and $V(z)$
plots. To understand the information given by these plots we start
the discussion of the $J(z)$ plots. Because these magnetic moment
values are very near to the zeros of the current we take this
property to claim that the sought free energy minimum must be such
that its $J(z)$ integrated along the z-axis will add to a zero
current. Fig.~\ref{vpj_z2_t08}(1a) shows that only the $L=1$ curve
qualifies for a free energy minimum since it is the only one to
cross the z-axis thus yielding positive current for $z<0$ and
negative for $z>0$. According to this criterion all others $L$ lines
can be dismissed as they can not be zero current states since they
do not cross the z-axis. This reasoning also provides the key to
select the free energy minimum in Figs.~\ref{vpj_z2_t08}(2a), (3a)
and (4a). The present criterion uniquely selects the $L=$ $7$, $12$
and $20$ states among the other ones drawn in these figures,
respectively, as the only possible free energy minimum because only
them fit as zero current states. For the short cylinder the
superconducting density $|\psi(z)|^2$, and the potential $V(z)$ also
provide features able to uniquely define the sought $L$ lines. For
the free energy minimum states the density is nearly constant
according to Figs.~\ref{vpj_z2_t08}(1b), (2b), (3b) and (4b).
Similarly the potential approaches zero for these states as seen in
Figs.~\ref{vpj_z2_t08}(1c), (2c), (3c) and (4c).

Fig.\ref{vpj_z10_t08} shows these plots for the $Z_0=10.0\xi_0$
cylinder at the selected magnetic moment values called $(1)$, $(6)$,
$(7)$ and $(11)$. Fig.~\ref{vpj_z10_t08}(1a) shows that in the low
magnetic moment regime the $L$ state associated to the free energy
minimum can be judiciously selected by the zero current condition
and this is the $L=1$ state. The same does not hold for the
intermediate magnetic moment values  $(6)$ and $(7)$ where many
lines qualify at least at the naked eye level precision, which means
that many lines cross the z-axis near the center of the cylinder
having positive current for $z<0$ and negative for $z>0$. For the
high magnetic moment limit the situation is similar although the
current vanishes near the upper edge of the cylinder, $z \approx 5$,
because of the proximity to the strong external current loop. The
dramatic changes that the superconducting state undergoes by
increasing the magnetic moment are witnessed  in the superconducting
density $|\psi(z)|^2$, and the potential $V(z)$.
Figs.~\ref{vpj_z10_t08}(1b), (2b), (3b) and (4b) show that density
is mostly concentrated at the top edge of the cylinder for
$\mu/\mu_0=12.75$ which falls in the low magnetic moment regime. For
the intermediate values of $\mu/\mu_0=85.64$ and $129.02$ we find
$L$ states concentrated at both edges to finally find that in the
high magnetic moment regime, represented by $\mu/\mu_0=264.97$, the
density is mostly concentrated at the bottom. Indeed
Figs.~\ref{vpj_z10_t08}(1c), (2c), (3c) and (4c) show that the
potential undergoes dramatic changes from the low to the high
magnetic moment regime. In the former case ($\mu/\mu_0=85.64$) it
undergoes mild changes from one edge to the other as it varies from
$0$ to $1$ in reduced units, while for the latter
case($\mu/\mu_0=264.97$) it varies from $0$ to $23$, being very
intense in the top edge. These plots explain the reasons for the
twofold regimes present in the LP oscillation near the persistent
current loop.

\section{Conclusion}\label{conclusion}

We have considered the Little-Parks oscillations on a thin-walled
superconducting cylinder with a persistent (magnetic moment) current
loop set on its top. This mesoscopic system is shown here to have
two regimes of temperature oscillations not found in the original
Little-Parks system because instead of a constant applied magnetic
field there is  an inhomogeneous magnetic field created in space by
the persistent current loop. We observe that this inhomogeneity
grows in relevance according to the cylinder height, as expected. In
order to unveil the new features we have considered a short and a
tall cylinder, and report new results in both cases. For the short
cylinder we report new analytical expressions for the magnetic
moments associated to the free energy minima, to the zeros of the
current, and to the maximum attainable temperature (Eqs.(\ref{muf}),
(\ref{muc}) and (\ref{mut})). For instance, we find that the current
in the thin-walled cylinder vanishes at a magnetic moment greater
than that of the free energy minimum ($\mu_c \ge \mu_f$). For the
tall cylinder we find novel features, perhaps best captured by
Fig.~\ref{curr_z10_t08t09}. This figure shows two distinct current
regimes for the tall cylinder, set at low and high $L$,
respectively, associated to distinct slopes with a clear transition
$L$ line that separates them. The existence of these two regimes,
and the transition that separates them, are within the realm of
observable experimental  measurements.

\section{Acknowledgements}
\label{acknowledgements}
This work is supported by the brazilian agencies CNPq and Facepe.

\end{document}